# Probing band topology in ABAB and ABBA stacked twisted double bilayer graphene

Jundong Zhu[1,2,†], Le Liu[1,2,†], Yalong Yuan[1,2,†], Jinwei Dong[1,2], Yanbang Chu[1,2], Luojun Du[1,2], Kenji Watanabe[4], Takashi Taniguchi[5], Jianpeng Liu[6], Quansheng Wu[1,2], Dongxia Shi[1,2,3], Wei Yang[1,2,3*] & Guangyu Zhang[1,2,3*]

[1] *Beijing National Laboratory for Condensed Matter Physics and Institute of Physics, Chinese Academy of Sciences, Beijing 100190, China*

[2] *School of Physical Sciences, University of Chinese Academy of Sciences, Beijing, 100190, China*

[3] *Songshan Lake Materials Laboratory, Dongguan 523808, China*

[4] *Research Center for Functional Materials, National Institute for Materials Science, 1-1 Namiki, Tsukuba 305-0044, Japan*

[5] *International Center for Materials Nanoarchitectonics, National Institute for Materials Science, 1-1 Namiki, Tsukuba 305-0044, Japan*

[6] *School of Physical Sciences and Technology, ShanghaiTech University, Shanghai 200031, China*

\* Corresponding authors. Email: wei.yang@iphy.ac.cn; gyzhang@iphy.ac.cn

(Dated 18$^{\text{th}}$ Sept. 2024)

## Abstract

Twisted graphene moiré superlattice has been demonstrated as an exotic platform for investigating correlated states and nontrivial topology. Among the moiré family, twisted double bilayer graphene (TDBG) is a tunable flat band system expected to show stacking-dependent topological properties. However, electron correlations and the band topology are usually intertwined in the flat band limit, rendering the unique topological property due to stacking still elusive. Focusing on a large-angle TDBG with weak electron correlations, here we probe the Landau level (LL) spectra in two differently stacked TDBG, i.e. ABBA- and ABAB-TDBG, to unveil their distinct topological properties. For ABBA-TDBG, we observe non-trivial topology at zero electric displacement filed D = 0, evident from both the emergence of Chern bands from half fillings $v = \pm 2$ and the closure of gap at CNP above a critical magnetic field. For ABAB-TDBG, by contrast, we find that the moiré band is topologically trivial at D = 0, supported by the absence of LLs from half fillings and the persistence of the gap at CNP above the critical magnetic fields. In addition, we also observe an evolution of the trivial-to-nontrivial topological transition at D ≠ 0, confirmed by the emerged Landau fans originating from quarter filling $v = 1$. Our result demonstrates, for the first time, the unique stacking-dependent topology in TDBG, offering a promising avenue for future investigations on topological states in correlated systems.

## Introduction

Twisted moiré superlattice has been demonstrated as an exotic platform for investigating correlated states and nontrivial band topology[1-24]. The nontrivial topology of the moiré bands is encoded in the nonzero Chern numbers (C) that can be measured from the quantized Hall resistance and the Streda formula C = (h/e)dn/dB[25]. In twisted bilayer graphene, electron correlations and band topology are intertwined, and spontaneous symmetry breakings would lead to exotic phases, such as quantized anomalous Hall effects at zero magnetic fields[4,5] and topological bands with high Chern numbers in a sequential filling fashion at finite magnetic fields[10-14]. In addition, stacking orders and the electric displacement fields (D) play a pivotal role in generating nontrivial band topology in graphene multilayers[6,26,27]. In particular, the fractional quantized anomalous Hall effects have been observed recently in pentalayer rhombohedral graphene[28].

Among the moiré family, twisted double bilayer graphene (TDBG) features a flat-band system tunable with D[17-21]. Recently, TDBG has been observed hosting spin ferromagnetism and various first-order quantum phase transitions[29], spin-valley competition[30] and valley-polarized phases[30,31], anomalous quantum oscillations in field-induced correlated insulators[32], and proximity-induced superconductivity[33]. TDBG is also expected to be a chirality-tunable topological system[34-37]. Generally, two stacking orders exist in TDBG, i.e., ABAB and ABBA. While both ABAB- and ABBA-TDBG share nearly identical band structures[34], their different symmetries ($C_{2y}$ vs. $C_{2x}$) yield distinct topological properties[34-39]. For example, the moiré bands in ABBA-TDBG have nonzero Chern numbers at different valleys, while these nontrivial topologies are absent in ABAB-TDBG[39,40]. Theoretical calculations also suggest that the different topologies should result in different Landau level (LL) spectra and the distinct magnetic-field dependence of the gap at the charge neutral point (CNP)[36,41]. It is worth noting that symmetry-breaking Chern insulators with nontrivial topology are observed for both stacking configurations at a finite displacement field[30,42]. However, the intertwining of strong electron correlation on band topology renders the unique topological properties of ABBA-TDBG still elusive.

In this work, we unveil the topological properties of ABBA-TDBG from the Landau Level fan diagram and the magnetic field dependence of the gap at CNP. This is achieved by focusing on TDBG with a large twist angle > 1.6°, to avoid flat band conditions for strong correlations and spontaneous symmetry breakings. For the first time, we demonstrate that the moiré bands in ABBA-TDBG are topologically nontrivial, evident from the emergence of LLs at half fillings $v = 2$ and the closure of the gap at high magnetic fields. In contrast, these features are absent in ABAB-TDBG. It is also worth mentioning that we observe the emergence of the symmetry-breaking LLs emanating from $v = 1$ in one ABAB-TDBG device at D ≠ 0. These observations demonstrate the intimate connection between the stacking symmetry and the nontrivial topology in a field tunable moiré system.

## Results and Discussions

The dual-gate TDBG devices (Figure 1a) for transport measurements are fabricated via the cut-and-stack technique[43]. As illustrated in Fig. 1b and 1c, ABAB- (ABBA-) TDBG is formed by stacking the two Bernal (i.e. AB) bilayer graphene sheets with a twist angle of $\theta$ (60°+$\theta$). Here, $\theta$ is around 1.8° to avoid spontaneous symmetry breakings and correlated states at zero magnetic fields. In the dual gate geometry, the carrier density (n) and D are defined as $n = (C_b V_b + C_t V_t)/e$ and $D = (C_b V_b - C_t V_t)/2\varepsilon_0$, where $C_b$ ($C_t$) is the geometrical capacitance per area for the bottom (top) gate, $e$ is the electron charge, and $\varepsilon_0$ is vacuum permittivity. Fig.1d shows typical band structures for both ABBA- and ABAB-TDBG at $\theta$ =1.8° based on the continuum model (see details in the Methods). The two stackings share the same band structure; ABBA-TDBG is topologically nontrivial with valley Chern number $C_v$ = -1 and 3 for conduction and valence bands, while ABAB-TDBG is topologically trivial with $C_v$ = 0. These results agree well with previous simulations[34,37]. Fig. 1e shows two transfer curves for ABBA-TDBG (pink) and ABAB-TDBG (grey) devices at D = 0, where the resistance peaks at CNP and those away are attributed to gap openings at CNP ($v = 0$) and at the full fillings ($v = 4$), respectively. Twist angles of $\theta = 1.73°$ and $1.87°$ are obtained for the ABBA-TDBG and the ABAB-TDBG devices from the carrier density at v = 4. Here, $v$ is the moiré filling factor $v = 4n/n_s$ with $n_s = 4/A \approx 8\theta/(\sqrt{3}a^2)$, where $A$ is the area of a moiré unit cell and $a$ is the lattice constant of graphene. Fig. 1f and 1g are the corresponding color maps of longitudinal resistance $R_{xx}$ as a function of $v$ and D for the ABBA- and ABAB- TDBG, respectively, sharing a similar phase diagram. In particular, the states around $v = 2$ are good metals with low resistance varying from tens to hundreds Ω, regardless of D. The observations agree with the absence of strong correlation and spontaneous symmetry-breaking states in the weakly coupled regime previously studied[44-48]. In short, these almost identical transport features at zero magnetic field for both devices agree well with their similar band structures, regardless of the stacking order.

To explore the band topology difference between the two stacking orders, we then measure the Landau fan diagrams at D = 0. We first focus on an ABBA-TDBG device at $\theta = 1.73°$. Fig. 2a and 2b show the corresponding Landau fan diagrams of longitudinal resistance $R_{xx}$ and Hall resistance $R_{xy}$, respectively. A series of LLs can be linearly traced by following the Diophantine equation $n/n_0 = C\phi/\phi_0 + s$, where C is the Chern number, $\phi$ is the magnetic flux in the moiré unit cell, $\phi_0$ = h/e is the quantum flux, and s is the Bloch band filling index. The trajectories of LLs in Fig. 2a and 2b are summarized in Fig. 2c. The Landau fans emanating from the full fillings at $v = \pm 4$ are delineated in black, and those from CNP are depicted in blue. Moreover, we could identify LLs fanning out of half fillings at $v = -2$ and 2, represented by red and pink lines, respectively. At $v = 2$, LLs with Chern numbers varying from -6 to -20 and degeneracy of 2 are observed, and they are confined in the phase diagram between the main LLs with C = 6 and those with C = 12 at CNP. At $v = -2$, LLs with Chern numbers varying from 22 to 32 and degeneracy of 2 are observed, and they are confined in the phase diagram between the main LLs with C = 0 and those with C = 6 at CNP. These two sets of LLs from half fillings intersect at the LL with C = 6 from CNP at B > ~5 T, as shown by the red and the pink solid lines in Fig. 2c. The two-fold degeneracy of LLs from half fillings agrees well with the previous theoretical calculations[36], indicating the distinct non-zero orbital magnetic moments of the valleys in the ABBA-TDBG that couples to the out-of-plane magnetic fields.

We also analyze the Landau fan diagram in the ABAB-TDBG. Fig. 2d-f show the results from a device at $\theta$ = 1.81°. We can see that LLs with the degeneracy of 4 fan out from both CNP at $v = 0$ and the full fillings at $v = \pm 4$ when magnetic fields are low, e.g. B < 4 T. While under high magnetic fields, e.g. B > 5 T, LLs with lifted degeneracy of 2, especially for the holes, are observed. Note that, similar to the ABBA-TDBG at $\theta$ = 1.73°, this device is also of high quality, evident from the low resistivity and the high mobility (Supplementary information Fig. S1). For another ABAB-TDBG device at $\theta$ = 1.87° that shows a relatively lower quality (Supplementary information Fig. S2), LLs with degeneracy of 4 are fanning out from CNP at $v = 0$ and the full fillings at $v = \pm 4$. For both ABAB-TDBG devices, regardless of their qualities, there are no obvious LL features fanning out from half fillings. The absence of LLs from half fillings, as well as the four-fold degeneracy of the LLs, agree well with the previously predicted trivial topology in ABAB-TDBG[36].

Besides, we also observe distinct responses of the resistivity at CNP to the magnetic field for the two different stackings. Fig. 3a shows the resistance plots at CNP against the magnetic field. For ABBA-TDBG (pink), it evolves from a good insulator with $R_{xx} > 20$ kΩ at B ≤ 6T to a metal with $R_{xx} < 1.2$ kΩ at B > 7 T. For ABAB-TDBG (cyan), by contrast, it preserves as a good insulator with $R_{xx} > 20$ kΩ. Based on the temperature dependence measurements, as shown in Fig. 3c and 3d, we can extract the thermal activation gap Δ by the Arrhenius equation $R_{xx} \propto \exp(\Delta/2k_BT)$, where $k_B$ is the Boltzmann constant. The extracted dependence of Δ on B is shown in Fig. 3b. For the ABBA-TDBG, the gap experiences a gradual increase with B till reaching a magnetic flux of $\phi/\phi_0 \sim 0.02$. Then, the gap undergoes a gradual reduction as B further increases. Eventually, the gap is fully closed at a magnetic flux of $\phi/\phi_0 \sim 0.1$. By contrast, for the ABAB-TDBG, the gap remains open at CNP and shows weak dependence on B (Fig. 3b and Supplementary information Fig. S3).

The magnetic field dependence of the gap at CNP reveals the distinct valley topological properties between ABBA- and ABAB-TDBG. At a finite magnetic field, moiré subbands are subjected to both the spin and orbital Zeeman effect. For ABBA-TDBG, the gap changes from ~8.2 meV at B = 0 T and ~11.67 meV at B = 1.5 T to ~0 meV at B = 7 T. Considering the small spin Landé g factor of g = 1, the contribution due to the spin Zeeman effect is smaller than 0.4 meV for B < 6 T thus negligible. The gap closure is thus attributed mainly to the orbital Zeeman effect as $\Delta \sim 2g\mu_B B$, where $\mu_B$ is the Bohr magneton. We then obtain a large g factor of g ~ 18 by fitting the linear regime for the magnetic field varying from B = 1.5 T to B = 7 T (Fig. 3b). Such a large orbital Zeeman effect agrees well with previous theory predictions of large valley contrast orbital magnetism[34-38]. For ABAB-TDBG, it changes from ~10.2 meV at B = 0 T and ~14.7 meV at B = 3 T to ~11.5 meV at B = 7 T. Similarly, we obtain a g factor of g ~7 by fitting the linear decreasing regime from B = 3 T to B = 7 T. If we take the moiré conduction band as trivial with $g_c = 0$[36], we could get a large $g_v \sim 14$ for the valence band, comparable to that in ABBA-TDBG. These results provide a quantitative agreement between the experiments and theory[36], further demonstrating the unique topology in ABBA-TDBG. Note that the gap also scales linearly at small magnetic fields for both ABBA- and ABAB-TDBG. This anomaly might indicate a complicated band structure and topology in magnetic fields.

Last but not least, the topological properties of TDBG can also be modulated by applying external electrical fields to break the $C_2$ symmetry[37-39]. Here, we demonstrate a topological phase transition in an ABAB-TDBG device at θ = 1.81° and investigate the influence of D on their topological properties. At zero displacement field D = 0, Landau fans originating from CNP are observable with LLs Chern numbers ($C_{LL}$) of 4, 8, 12, and so on, indicating a degeneracy of 4 (see Fig. 4a). At an elevated electric field of D = -0.369 V/nm, the first three LLs from CNP evolves to the ones with $C_{LL}$ = 2, 5, 8 (see Fig. 4b). More importantly, we observe the emergence of LLs with $C_{LL}$ = -5, -6, -7, and -8 from $v$ = 1 in Hall measurements (as depicted in Fig. 4e), residing in the phase space between the main LLs $C_{LL}$ = 2 and 5 from CNP at B > 6T. In particular, these new symmetry-breaking LLs show a sign change of the Hall resistance in Fig. 4e, where the negative Chern numbers are in stark contrast to the nearby red positive ones from CNP, as well as those at D = 0 V/nm in Fig. 4d. Signatures of full symmetry broken LLs are also observed in the $R_{xx}$ mapping with $C_{LL}$ = 1, 0, -1 emanating from $v$ = 1 (as depicted in Fig. 4b). Upon reaching an electric field strength of D = -0.44 V/nm, the Landau fans from CNP develops into LLs with $C_{LL}$ = 2, 3, 6, and others; by comparison, LLs with $C_{LL}$ = -1, -2, -3, and -4 are developed at $v$ = 1. Generally, these findings indicate that the external electric field breaks the $C_{2x}$ symmetry of the ABAB-TDBG device and leads to nontrivial topology revealed in the new LLs generated at quarter fillings, consistent with the intimate relationship between the topology and symmetry breaking from previous calculations[34-36].

## Conclusions

We have unveiled the unique topology originating from stacking order in a weakly moiré system facilitated by twisting two bilayer graphene sheets with relatively large twist angles to minimize the correlation effects on band topology. The topological properties from the LL spectra and magnetic field dependence of the gap at CNP for the two stackings, i.e. ABBA and ABAB configurations, are systematically probed. The moiré bands from ABBA-TDBG are found to be topologically nontrivial at zero electric displacement field, supported by the observations of the new LLs developed from half fillings and the gap closure at CNP under high magnetic fields. In addition, for the moiré band in ABAB-TDBG, we observe a topological phase transition from trivial bands at zero electric displacement field to nontrivial bands at a finite electric displacement field, highlighting the symmetry breaking of $C_{2x}$. Our results demonstrate rich topological phases and the importance of stacking order in TDBG, and it might shed light on exotic topological phases in the strongly correlated regime in TDBG and other moiré systems as well.

## References

1. Cao Y, Fatemi V, Demir A, Fang S, Tomarken SL, Luo JY, Sanchez-Yamagishi JD, Watanabe K, Taniguchi T, Kaxiras E, Ashoori RC, Jarillo-Herrero P. Correlated insulator behaviour at half-filling in magic-angle graphene superlattices. *Nature* **556**, 80-84 (2018).
2. Cao Y, Fatemi V, Fang S, Watanabe K, Taniguchi T, Kaxiras E, Jarillo-Herrero P. Unconventional superconductivity in magic-angle graphene superlattices. *Nature* **556**, 43-50 (2018).
3. Lu X, Stepanov P, Yang W, Xie M, Aamir MA, Das I, Urgell C, Watanabe K, Taniguchi T, Zhang G, Bachtold A, MacDonald AH, Efetov DK. Superconductors, orbital magnets and correlated states in magic-angle bilayer graphene. *Nature* **574**, 653-657 (2019).
4. Sharpe AL, Fox EJ, Barnard AW, Finney J, Watanabe K, Taniguchi T, Kastner MA, Goldhaber-Gordon D. Emergent ferromagnetism near three-quarters filling in twisted bilayer graphene. *Science* **365**, 605-608 (2019).
5. Serlin M, Tschirhart CL, Polshyn H, Zhang Y, Zhu J, Watanabe K, Taniguchi T, Balents L, Young AF. Intrinsic quantized anomalous hall effect in a moiré heterostructure. *Science* **367**, 900-903 (2020).
6. Chen G, Sharpe AL, Fox EJ, Zhang YH, Wang S, Jiang L, Lyu B, Li H, Watanabe K, Taniguchi T, Shi Z, Senthil T, Goldhaber-Gordon D, Zhang Y, Wang F. Tunable correlated Chern insulator and ferromagnetism in a moiré superlattice. *Nature* **579**, 56-61 (2020).
7. Nuckolls KP, Oh M, Wong D, Lian B, Watanabe K, Taniguchi T, Bernevig BA, Yazdani A. Strongly correlated Chern insulators in magic-angle twisted bilayer graphene. *Nature* **588**, 610-615 (2020).
8. Choi Y, Kim H, Peng Y, Thomson A, Lewandowski C, Polski R, Zhang Y, Arora HS, Watanabe K, Taniguchi T, Alicea J, Nadj-Perge S. Correlation-driven topological phases in magic-angle twisted bilayer graphene. *Nature* **589**, 536-541 (2021).
9. Saito Y, Ge J, Rademaker L, Watanabe K, Taniguchi T, Abanin DA, Young AF. Hofstadter subband ferromagnetism and symmetry-broken Chern insulators in twisted bilayer graphene. *Nature Physics* **17**, 478-481 (2021).
10. Wu S, Zhang Z, Watanabe K, Taniguchi T, Andrei EY. Chern insulators, van hove singularities and topological flat bands in magic-angle twisted bilayer graphene. *Nature Materials* **20**, 488-494 (2021).
11. Das I, Lu X, Herzog-Arbeitman J, Song Z-D, Watanabe K, Taniguchi T, Bernevig BA, Efetov DK. Symmetry-broken Chern insulators and rashba-like landau-level crossings in magic-angle bilayer graphene. *Nature Physics* **17**, 710-714 (2021).
12. Park JM, Cao Y, Watanabe K, Taniguchi T, Jarillo-Herrero P. Flavour hund's coupling, Chern gaps and charge diffusivity in moiré graphene. *Nature* **592**, 43-48 (2021).
13. Xie Y, Pierce AT, Park JM, Parker DE, Khalaf E, Ledwith P, Cao Y, Lee SH, Chen S, Forrester PR, Watanabe K, Taniguchi T, Vishwanath A, Jarillo-Herrero P, Yacoby A. Fractional Chern insulators in magic-angle twisted bilayer graphene. *Nature* **600**, 439-443 (2021).
14. Shen C, Ying J, Liu L, Liu J, Li N, Wang S, Tang J, Zhao Y, Chu Y, Watanabe K, Taniguchi T, Yang R, Shi D, Qu F, Lu L, Yang W, Zhang G. Emergence of Chern insulating states in non-magic angle twisted bilayer graphene. *Chinese Physics Letters* **38**, 047301 (2021).
15. Cai J, Anderson E, Wang C, Zhang X, Liu X, Holtzmann W, Zhang Y, Fan F, Taniguchi T, Watanabe K, Ran Y, Cao T, Fu L, Xiao D, Yao W, Xu X. Signatures of fractional quantum anomalous hall states in twisted $MoTe_2$. *Nature* **622**, 63-68 (2023).
16. Xu F, Sun Z, Jia T, Liu C, Xu C, Li C, Gu Y, Watanabe K, Taniguchi T, Tong B, Jia J, Shi Z, Jiang S, Zhang Y, Liu X, Li T. Observation of integer and fractional quantum anomalous hall effects in twisted bilayer $MoTe_2$. *Physical Review X* **13**, 031037 (2023).
17. Burg GW, Zhu J, Taniguchi T, Watanabe K, MacDonald AH, Tutuc E. Correlated insulating states in twisted double bilayer graphene. *Phys Rev Lett* **123**, 197702 (2019).
18. Shen C, Chu Y, Wu Q, Li N, Wang S, Zhao Y, Tang J, Liu J, Tian J, Watanabe K, Taniguchi T, Yang R, Meng ZY,

|   | Shi D, Yazyev OV, Zhang G. Correlated states in twisted double bilayer graphene. *Nature Physics* **16**, 520-525 (2020). |
|---|---|
| 19 | Cao Y, Rodan-Legrain D, Rubies-Bigorda O, Park JM, Watanabe K, Taniguchi T, Jarillo-Herrero P. Tunable correlated states and spin-polarized phases in twisted bilayer-bilayer graphene. *Nature* **583**, 215-220 (2020). |
| 20 | Liu X, Hao Z, Khalaf E, Lee JY, Ronen Y, Yoo H, Haei Najafabadi D, Watanabe K, Taniguchi T, Vishwanath A, Kim P. Tunable spin-polarized correlated states in twisted double bilayer graphene. *Nature* **583**, 221-225 (2020). |
| 21 | Wang Y, Herzog-Arbeitman J, Burg GW, Zhu J, Watanabe K, Taniguchi T, MacDonald AH, Bernevig BA, Tutuc E. Bulk and edge properties of twisted double bilayer graphene. *Nature Physics* **18**, 48-53 (2021). |
| 22 | Polshyn H, Zhu J, Kumar MA, Zhang Y, Yang F, Tschirhart CL, Serlin M, Watanabe K, Taniguchi T, MacDonald AH, Young AF. Electrical switching of magnetic order in an orbital Chern insulator. *Nature* **588**, 66-70 (2020). |
| 23 | Chen S, He M, Zhang Y-H, Hsieh V, Fei Z, Watanabe K, Taniguchi T, Cobden DH, Xu X, Dean CR, Yankowitz M. Electrically tunable correlated and topological states in twisted monolayer–bilayer graphene. *Nature Physics* **17**, 374-380 (2020). |
| 24 | Xu S, Al Ezzi MM, Balakrishnan N, Garcia-Ruiz A, Tsim B, Mullan C, Barrier J, Xin N, Piot BA, Taniguchi T, Watanabe K, Carvalho A, Mishchenko A, Geim AK, Fal'ko VI, Adam S, Neto AHC, Novoselov KS, Shi Y. Tunable van hove singularities and correlated states in twisted monolayer–bilayer graphene. *Nature Physics* **17**, 619-626 (2021). |
| 25 | Streda P. Theory of quantised hall conductivity in two dimensions. *Journal of Physics C: Solid State Physics* **15**, L717-L721 (1982). |
| 26 | Sha Y, Zheng J, Liu K, Du H, Watanabe K, Taniguchi T, Jia J, Shi Z, Zhong R, Chen G. Observation of a Chern insulator in crystalline abca-tetralayer graphene with spin-orbit coupling. *Science* **384**, 414-419 (2024). |
| 27 | Han T, Lu Z, Yao Y, Yang J, Seo J, Yoon C, Watanabe K, Taniguchi T, Fu L, Zhang F, Ju L. Large quantum anomalous hall effect in spin-orbit proximitized rhombohedral graphene. *Science* **384**, 647-651 (2024). |
| 28 | Lu Z, Han T, Yao Y, Reddy AP, Yang J, Seo J, Watanabe K, Taniguchi T, Fu L, Ju L. Fractional quantum anomalous hall effect in multilayer graphene. *Nature* **626**, 759-764 (2024). |
| 29 | Liu L, Lu X, Chu Y, Yang G, Yuan Y, Wu F, Ji Y, Tian J, Watanabe K, Taniguchi T, Du L, Shi D, Liu J, Shen J, Lu L, Yang W, Zhang G. Observation of first-order quantum phase transitions and ferromagnetism in twisted double bilayer graphene. *Physical Review X* **13**, 031015 (2023). |
| 30 | Liu L, Zhang S, Chu Y, Shen C, Huang Y, Yuan Y, Tian J, Tang J, Ji Y, Yang R, Watanabe K, Taniguchi T, Shi D, Liu J, Yang W, Zhang G. Isospin competitions and valley polarized correlated insulators in twisted double bilayer graphene. *Nat Commun* **13**, 3292 (2022). |
| 31 | Kuiri M, Coleman C, Gao Z, Vishnuradhan A, Watanabe K, Taniguchi T, Zhu J, MacDonald AH, Folk J. Spontaneous time-reversal symmetry breaking in twisted double bilayer graphene. *Nature Communications* **13**, 6468 (2022). |
| 32 | Liu L, Chu Y, Yang G, Yuan Y, Wu F, Ji Y, Tian J, Yang R, Watanabe K, Taniguchi T, Long G, Shi D, Liu J, Shen J, Lu L, Yang W, Zhang G. Quantum oscillations in field-induced correlated insulators of a moiré superlattice. *Sci Bull (Beijing)* **68**, 1127-1133 (2023). |
| 33 | Su R, Kuiri M, Watanabe K, Taniguchi T, Folk J. Superconductivity in twisted double bilayer graphene stabilized by $WSe_2$. *Nature Materials* **22**, 1332-1337 (2023). |
| 34 | Koshino M. Band structure and topological properties of twisted double bilayer graphene. *Physical Review B* **99**, 235406 (2019). |
| 35 | Crosse JA, Nakatsuji N, Koshino M, Moon P. Hofstadter butterfly and the quantum hall effect in twisted double bilayer graphene. *Physical Review B* **102**, 035421 (2020). |
| 36 | Wu Q, Liu J, Guan Y, Yazyev OV. Landau levels as a probe for band topology in graphene moiré superlattices. *Phys Rev Lett* **126**, 056401 (2021). |
| 37 | Chebrolu NR, Chittari BL, Jung J. Flat bands in twisted double bilayer graphene. *Physical Review B* **99**, 235417 (2019). |
| 38 | Liu J, Ma Z, Gao J, Dai X. Quantum valley hall effect, orbital magnetism, and anomalous hall effect in twisted multilayer graphene systems. *Physical Review X* **9**, 031021 (2019). |
| 39 | Lee JY, Khalaf E, Liu S, Liu X, Hao Z, Kim P, Vishwanath A. Theory of correlated insulating behaviour and spin- |

triplet superconductivity in twisted double bilayer graphene. *Nat Commun* **10**, 5333 (2019).

40   Wu F, Das Sarma S. Ferromagnetism and superconductivity in twisted double bilayer graphene. *Physical Review B* **101**, 155149 (2020).

41   Hofstadter DR. Energy levels and wave functions of bloch electrons in rational and irrational magnetic fields. *Physical Review B* **14**, 2239-2249 (1976).

42   He MH, Cai JQ, Zhang YH, Liu Y, Li YH, Taniguchi T, Watanabe K, Cobden DH, Yankowitz M, Xu XD. Symmetry-broken Chern insulators in twisted double bilayer graphene. *Nano Letters* **23**, 11066-11072 (2023).

43   Kim K, Yankowitz M, Fallahazad B, Kang S, Movva HCP, Huang SQ, Larentis S, Corbet CM, Taniguchi T, Watanabe K, Banerjee SK, LeRoy BJ, Tutuc E. Van der waals heterostructures with high accuracy rotational alignment. *Nano Letters* **16**, 1989-1995 (2016).

44   de Vries FK, Zhu J, Portolés E, Zheng G, Masseroni M, Kurzmann A, Taniguchi T, Watanabe K, MacDonald AH, Ensslin K, Ihn T, Rickhaus P. Combined minivalley and layer control in twisted double bilayer graphene. *Physical Review Letters* **125**, 176801 (2020).

45   Rickhaus P, de Vries FK, Zhu J, Portoles E, Zheng G, Masseroni M, Kurzmann A, Taniguchi T, Watanabe K, MacDonald AH, Ihn T, Ensslin K. Correlated electron-hole state in twisted double-bilayer graphene. *Science* **373**, 1257-1260 (2021).

46   Tomic P, Rickhaus P, Garcia-Ruiz A, Zheng G, Portoles E, Fal'ko V, Watanabe K, Taniguchi T, Ensslin K, Ihn T, de Vries FK. Scattering between minivalleys in twisted double bilayer graphene. *Phys Rev Lett* **128**, 057702 (2022).

47   Chu Y, Liu L, Shen C, Tian J, Tang J, Zhao Y, Liu J, Yuan Y, Ji Y, Yang R, Watanabe K, Taniguchi T, Shi D, Wu F, Yang W, Zhang G. Temperature-linear resistivity in twisted double bilayer graphene. *Physical Review B* **106**, 035107 (2022).

48   Yuan Y, Liu L, Zhu J, Dong J, Chu Y, Wu F, Du L, Watanabe K, Taniguchi T, Shi D, Zhang G, Yang W. Interplay of landau quantization and interminivalley scatterings in a weakly coupled moiré superlattice. *Nano Letters* **24**, 6722-6729 (2024).

49   Wang L, Meric I, Huang PY, Gao Q, Gao Y, Tran H, Taniguchi T, Watanabe K, Campos LM, Muller DA, Guo J, Kim P, Hone J, Shepard KL, Dean CR. One-dimensional electrical contact to a two-dimensional material. *Science* **342**, 614-617 (2013).

## Method

**Device fabrications.** Both bilayer graphene and h-BN flakes (thickness: 15-40 nm) were exfoliated on $SiO_2$ (thickness: 285 nm) substrate. Before stacking, h-BN flakes were etched in $H_2$ plasma at 450 °C for 2 hours to clean the surface. Both the ABBA-TDBG and ABAB-TDBG samples were fabricated using the 'cut and stack' method. Firstly, one AB stacked bilayer graphene flake was cut into two pieces by a tungsten tip (tip radius < 1 μm). Secondly, a h-BN flake (the top capsulation layer) was picked up by a stamp of PC (Poly (Bisphenol A carbonate))/PDMS (polydimethylsiloxane)/glass slide at 90 °C. Thirdly, a piece of bilayer graphene was first picked up from the $SiO_2$ substrate at 80 °C by the h-BN flake, and the other piece was picked up subsequently with a rotation angle of 60° + $\theta$ ($\theta$) relative to the first piece, forming ABBA-TDBG (ABAB-TDBG). Finally, another h-BN flake (the bottom capsulation layer) and a few-layer graphite flake (the bottom gate electrode) were sequentially picked up at 100 °C. The stacked samples were placed on pre-cleaned Si(p++)/$SiO_2$ substrates and then rinsed in chloroform to remove residues. Afterward, the samples were fabricated into a dual-gate Hall bar device using electron beam lithography, electron beam evaporation, and $CHF_3/O_2$ plasma etching. Ti/Au

(3nm/30nm) was deposited as the top gate electrode and few-layer graphite (FLG) or heavily doped silicon were used as the back-gate electrode. Cr/Au (3nm/30nm) were deposited as the edge contacts[49].

**Electrical measurements.** The electrical measurements were carried out in helium-4 vapor flow cryostats (Janis) with a base temperature of 1.8 K and a maximum magnetic field of 9T. The transport data were measured by standard lock-in techniques with a current of 5-10 nA at a frequency of 30.9 Hz (Stanford Research SR830).

**Twist angle Estimation.** We calculate the carrier density $n$ and the electric displacement field $D$ according to the capacitances and transform the $R_{xx}(V_b, V_t)$ mapping to the $R_{xx}(n, D)$ mapping. In the dual gate geometry, the carrier density (n) and the displacement field $D$ are defined as $n = (C_b V_b + C_t V_t)/e$ and $D = (C_b V_b - C_t V_t)/2\varepsilon_0$, where $C_b$ ($C_t$) is the geometrical capacitance per area for bottom (top) gate, $e$ is the electron charge, and $\varepsilon_0$ is vacuum permittivity. According to the formula, $n_s = 4/A \approx 8\theta^2/(\sqrt{3}a^2)$, the twisted angle $\theta$ can be extracted from the location of the superlattice resistance peak in the $R_{xx}(n, D)$ mapping, which is further corrected by the Landau fan diagram.

**Band structure and Chern number calculations.** The energy band calculations are performed with a continuum model. Considering the two stacking orders, we calculate the band structures for both AB-AB and AB-BA configurations. The Hamiltonian of bilayer graphene is written in the basis ($A_1$, $B_1$, $A_2$, $B_2$, $A_3$, $B_3$). Here $A_i$ and $B_i$ denote the A and B sublattice sites of the $i$ layer. The parameters we used are ($\gamma_0$, $\gamma_1$, $\gamma_3$, $\gamma_4$) = (3100, 400, 320, 44) meV. Here $\gamma_0$ is the nearest intralayer hopping term; $\gamma_1$, $\gamma_3$ and $\gamma_4$ are the interlayer hopping terms between two adjacent layers. The coupling terms between two twisted layers are set to ($u_{AA}$, $u_{AB}$) = (80, 100) meV by considering the corrugation effect. In addition, we calculate the valley Chern number for both configurations. For ABBA-TDBG, the band is topologically nontrivial with the valley Chern number of conduction (valance) band $C_v = -1$ (3); for ABAB-TDBG, however, it is topologically trivial with $C_v = 0$ (0).

## Acknowledgments

We acknowledge support from the National Key Research and Development Program (Grant No. 2020YFA0309600, 2021YFA1202900), the National Natural Science Foundation of China (NSFC, Grant Nos. 12074413, 61888102), the Strategic Priority Research Program of CAS (Grant Nos. XDB33000000). K.W. and T.T. acknowledge support from the JSPS KAKENHI (Grant Numbers 20H00354, 21H05233 and 23H02052) and World Premier International Research Center Initiative (WPI), MEXT, Japan.

## Author contributions

W.Y. and G.Z. supervised the project; W.Y. designed the experiments; J.Z., L.L., and Y.Y. fabricated the devices and performed the magneto-transport measurements, assisted by J.D. and Y.C.; L.L. performed the calculations; K.W. and T.T. provided hexagonal boron nitride crystals; W.Y., J.Z., L.L., Y.Y., and G.Z. analyzed the data; W.Y., J.Z., L.L., and G.Z. wrote the paper with the input from all the authors.

## Competing interests

The authors declare no competing interests.

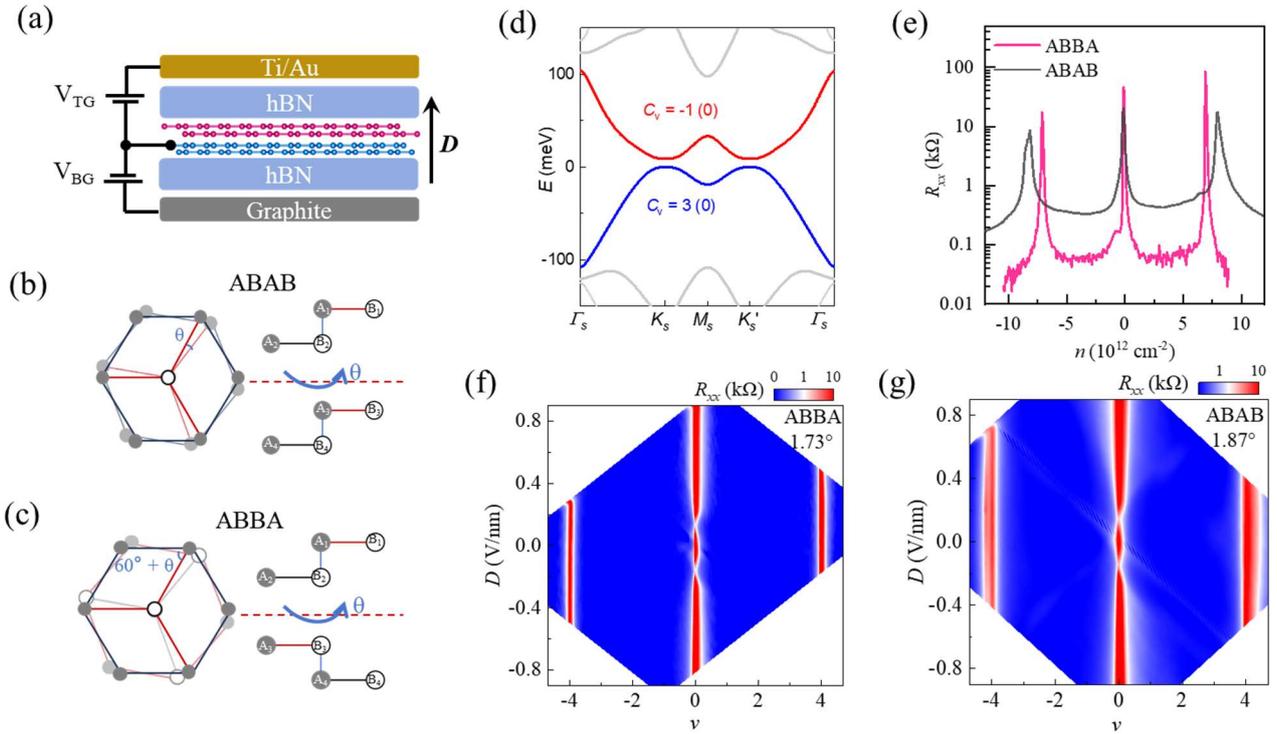

**Figure 1. TDBG devices with ABBA and ABAB stacking orders**. (**a**) Schematics of the dual-gated TDBG device. The arrow indicates a positive displacement electric field. (**b, c**) Schematics of AB-BA stacked and AB-AB stacked TDBG. (**d**) The calculated band structure of ABBA-TDBG (the same as ABAB-TDBG) at a twist angle of θ=1.8° from the continuum model. The ABBA-TDBG shows a nontrivial band topology with $C_v$ = -1 and 3 for electrons and holes, respectively, while that in ABAB-TDBG is topologically trivial with $C_v$ = 0 for both electrons and holes. (**e**) Typical transfer curves of TDBG devices at θ > 1.7° for both the ABAB (pink) and ABBA (grey) configurations at D = 0 and T = 1.8 K. (**f-g**) Color maps of $R_{xx}$(v, D) for the two devices in (**e**), where θ = 1.73° and 1.87° are obtained for ABBA-TDBG (**f**) and ABAB-TDBG (**g**), respectively.

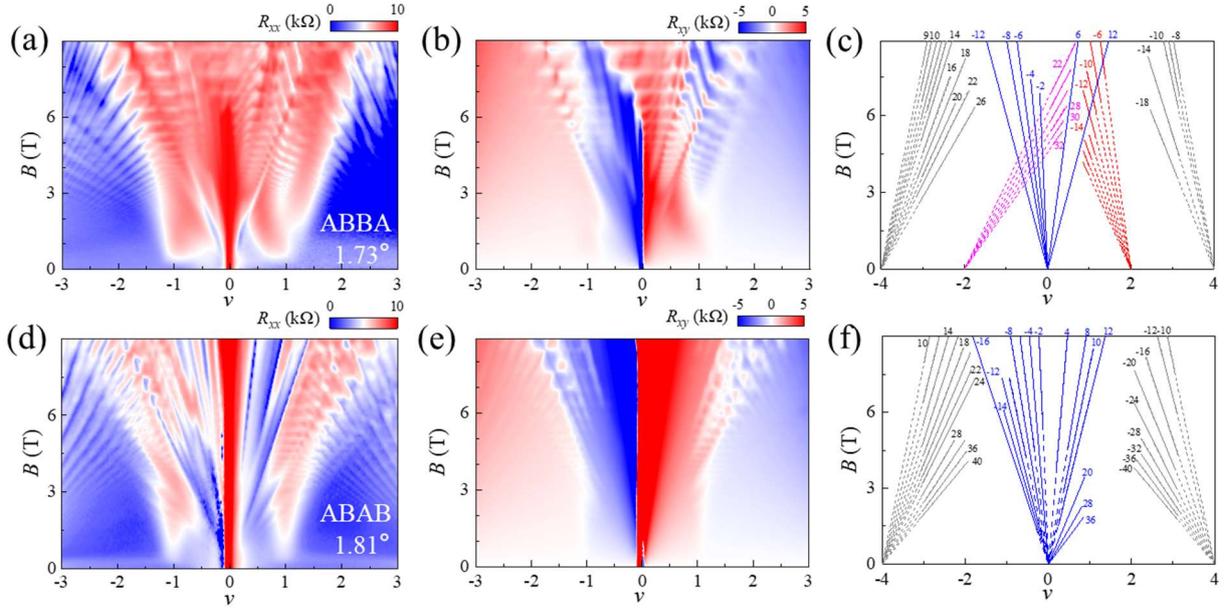

**Figure 2. Landau level spectra of typical TDBG devices with different stacking orders**. (**a-b**) Color maps of $R_{xx}$ (**a**) and $R_{xy}$ (**b**) as a function of moiré filling factor $v$ and magnetic field B for the ABBA-TDBG device at θ= 1.73°. (**c**) Wannier diagram obtained from Landau fan diagrams in (**a-b**). The LLs from half fillings v = -2 and 2 are marked in red, while those from CNP and full fillings are marked in blue and grey, respectively. (**d-e**) Landau fan diagram of $R_{xx}$ and $R_{xy}$ for the ABAB-TDBG device at θ = 1.81°. (**f**) Wannier diagram obtained from mappings in (**d-e**) shows the absence of the LLs fanning out from half fillings. All measures are at T = 1.8 K with D = 0.

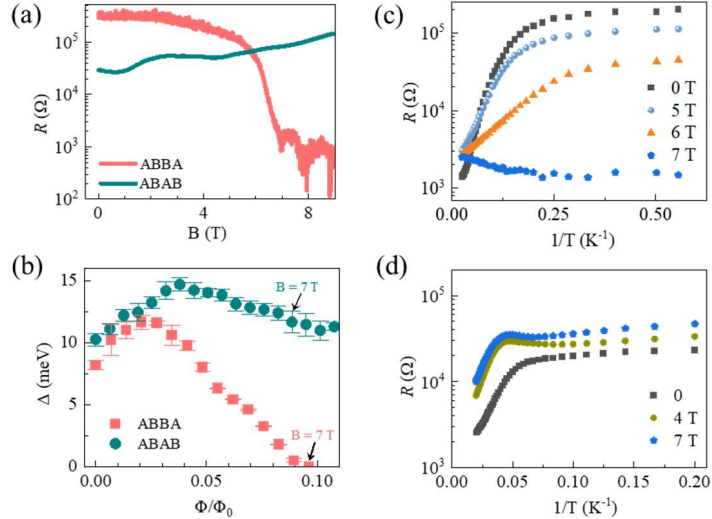

**Figure 3. Evolution of the gap at CNP under magnetic fields**. (**a**) Plots of $R_{xx}$ as a function of B, for both AB-BA TDBG (pink) and AB-AB TDBG (green) devices at T = 1.8 K. (**b**) Plots of thermal activation gaps at the CNP as a function of magnetic flux $\Phi/\Phi_0$ for the two devices. (**c**) Plots of $R_{xx}$ (1/T) at B = 0 T, 5 T, 6 T, and 7 T in ABBA-TDBG. (**d**) Plots of $R_{xx}$ (1/T) at B= 0T, 4T, and 7T in ABAB-TDBG.

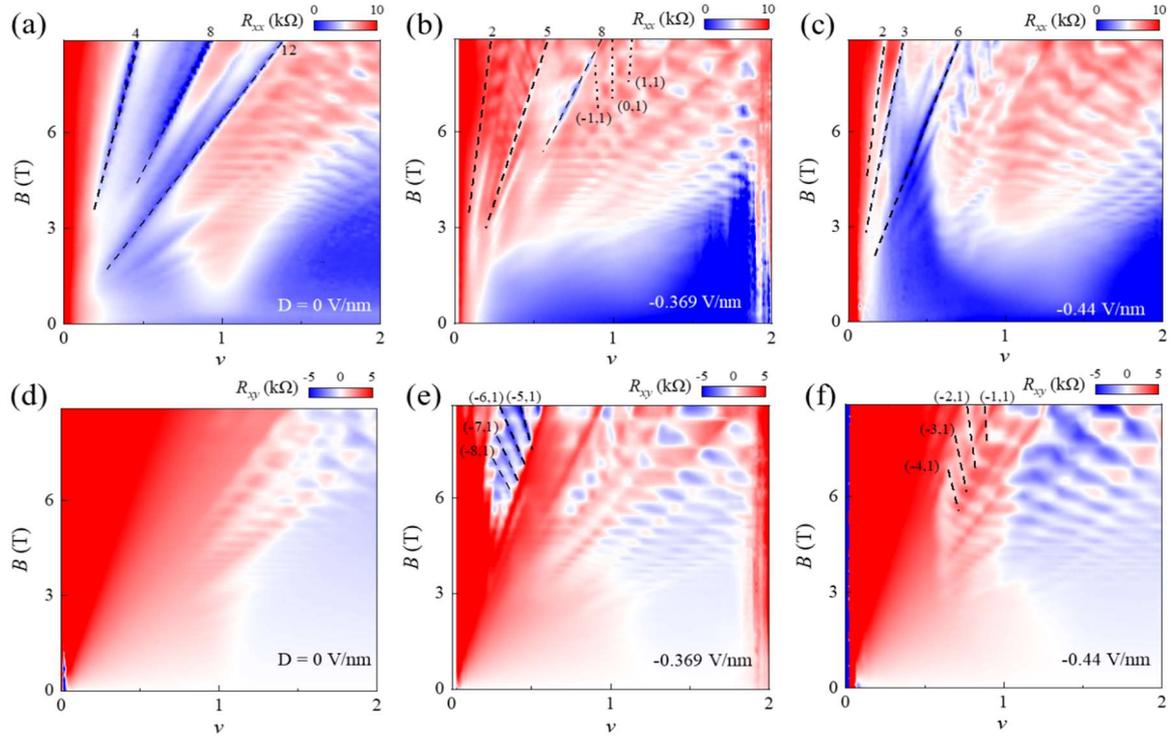

**Figure 4. Displacement field induced symmetry breakings in the Landau fan diagram for ABAB-TDBG at θ = 1.81°.** (**a-c**) Color maps of $R_{xx}$ at D = 0, -0.369, and -0.44 V/nm, respectively. (**d-f**) Color maps of $R_{xy}$ at D = 0, -0.369, and -0.44 V/nm, respectively. Those lines emanating from $v$ = 1 are labeled by their respective (C, 1) values, and those lines emanating from CNP are labeled by their respective C values.